\documentclass[11pt]{amsart}

\usepackage{amsmath,amsthm}
\usepackage{graphicx}
\usepackage{amsfonts}
\usepackage{amssymb}
\usepackage{natbib}

\theoremstyle{definition}
\newtheorem{example}{Example}

\title[Process, Population, and Sample]{Process, Population, and Sample: \\The Researcher's Interest}
\author{Charles W. Champ}
\thanks{Charles W. Champ is the lead author of this article.}
\email{cchamp@georgiasouthern.edu}
\address{Charles W. Champ, Department of Mathematical Sciences, Georgia Southern University,
Statesboro, Georgia, U.S.A.}

\author{Andrew V. Sills}
\thanks{Andrew Sills (asills@georgiasouthern.edu) is the corresponding author.}
\email{asills@georgiasouthern.edu}
\urladdr{http://home.dimacs.rutgers.edu/\~asills}
\address{Andrew V. Sills, Department of Mathematical Sciences, Georgia Southern University,
Statesboro and Savannah, Georgia, U.S.A.}

\begin{document}

\maketitle

\begin{abstract} A case is made that researchers are interested in
studying processes. Often the inferences they are interested in making are
about the process and its associated population. On other occasions, a
researcher may be interested in making an inference about the collection of
individuals the process has generated. We will call the statistical methods
employed by the researcher to make such inferences about the
process/population \textquotedblleft estimation methods.\textquotedblright\
The statistical methods used in making an inference about the collection of
individuals generated we call \textquotedblleft prediction
methods.\textquotedblright\ Methods for obtaining interval estimates of a
parameter and prediction intervals for a statistic are given. The analytical
and enumerative methods discussed in Deming (1953) are simply estimation and
prediction methods, respectively.
\end{abstract}

\section{Introduction}

Researchers draw inferences from data. \ Data is obtained from taking
measurements on individuals. \ Immediately, one is faced with the
fundamental question, \textquotedblleft Where do these individuals come
from?\textquotedblright\ \ The individuals are created by a repetative
process---natural, human made, or some combination thereof. \ Many authors
refer to the set of all individuals that were created by a process as the
\textquotedblleft population of interest,\textquotedblright\ but we regard
this view as incorrect, or at least incomplete. \ To begin to make our case,
we return to the thinking of R. A. Fisher.

Fisher believed statistics to be a branch of applied mathematics. \ He
stated in 
\citet[p. 1]{F58}%
: \textquotedblleft \lbrack t]he science of statistics is essentially a
branch of Applied Mathematics, and may be regarded as mathematics applied to
observational data. As in other mathematical studies, the same formula is
equally relevant to widely different groups of subject-matter. Consequently
the unity of the different applications had usually been overlooked, the
more naturally because the development of the underlying mathematical theory
had been much neglected.\textquotedblright\ It was not clear what he meant
by \textquotedblleft applied\textquotedblright\ mathematics, and indeed
there is no universally agreed upon definition of \textquotedblleft applied
mathematics.\textquotedblright\ Nonetheless, \textquotedblleft statistical
thinking\textquotedblright\ would then be just a special case of
\textquotedblleft mathematical thinking\textquotedblright\ according to
him.\ \cite{F58} further regarded statistics as \textquotedblleft (\textit{i}%
) the study of populations, (\textit{ii}) as the study of
variation,\textquotedblright\ and \textquotedblleft (\textit{iii}) as the
study of methods of the reduction of data.\textquotedblright

A variety of researchers and practitioners make use of statistical methods
to analyze their data. These applications include, among others,
environmental, pharmaceutical, medical, legal, biological, political, and
industrial data. The techniques used to analyze these data range from simple
graphical representations of data to rather involved formal statistical
methods. At the heart of these applications, is a \textit{process} and
associated \textit{population} of interest to the researcher. The complete
information about the process and population will never be available to the
researcher. The information the researcher will have available is always
partial information about the process and population in the form of a finite
subcollection of the population called a \textit{sample}. In this article,
we will discuss the meaning of a population and sample. Further, we will
examine and explicitly differentiate two statistical methods of inference, 
\textbf{estimation} and \textbf{prediction}. We will argue that the two
areas of statistical inference, called \textbf{analytical }and \textbf{%
enumerative} in \cite{D53} are, respectively, estimation and predication as
discussed in this article.

\section{Population and Sample}

An individual to be measured by a researcher is one that has been produced
by a repetative process. This process may be natural, human made, or a
combination of the two. The collection of individuals the process has
generated (actual individuals) or could have generated (conceptual
individuals) is the researcher's \textbf{population} of interest. \ We
assert that the collection of conceptual individuals must be not only an
infinite collection (as stated by Fisher, quoted below), but is inherently
an \textit{uncountably infinite} collection. \ Accordingly, the population
(as it contains both hypothetical individuals and a finite number of actual
individuals produced by the process) is also an uncountably infinite
collection of individuals. To see this, consider the following example:

\begin{example}
Consider the simple experiment of tossing a coin once, and observing whether
the coin lands heads up or tails up. The population consists of all possible
coin tosses. For simplicity, suppose that we have one person designated as
the coin tosser. Now the coin tosser can stand (or sit) in any of a range of
positions. Considering the floor to be a two-dimensional surface, there are
already uncountably infinitely many standing/sitting locations even within a
small area in the room. The height above the floor from which the coin is
tossed can also be an interval of values, and again, any interval of
positive length in the real numbers contains uncountably many points. There
are also slight variants in the angle at which the coin is launched, the
amount of force applied in the launch of the coin, and even variants in the
air currents through which the coin will travel once released. All of these
reinforce the idea of an uncountably infinite number of hypothetical
individuals (coin tosses) contained in the population.
\end{example}

Having named the collection of all actual and hypothetical individuals
associated with a given process as the population, we need a different name
for the (finite) collection of \textit{actual} individuals the process has
generated. \ This collection of actual individuals is a sample
representative of what the process can generate, so let us call it the 
\textbf{representative\ sample.} This representative sample is a finite
collection of some cardinality, say $N$. \ Again, many authors refer to the
representative sample as \textquotedblleft the
population,\textquotedblright\ but we believe this to be incorrect for a
number of reasons. \ This collection generated by the process is just 
\textit{one} of an \textit{uncountable collection of samples of size }$N$%
\textit{\ that the process could have generated}. One might refer to this
latter collection as the \textbf{population of representative samples}. The
uncertainty as to \textit{which} representative sample the process has
generated \textit{must be taken into account}, and will inevitably be
ignored if we view the representative sample as the \textquotedblleft
(finite) population\textquotedblright\ of interest. Note that the size $N$
of the representative sample is in fact a random variable. A subcollection
of the representative sample of size $n$, where $1\leq n\leq N$, selected by
the researcher is what we shall call the \textbf{researcher's sample}.\ The
sampling methods available to the researcher at best gives the researcher a
good chance of obtaining a sample that is representative of the
representative sample. The researcher's sample is a \textbf{census} if $n=N$.

It is not difficult to argue via examples that a researcher is interested in
making inferences about a process and the associated population \textit{or}
about the representative sample the process has generated. Most often
researchers are interested in making inferences about a process or
population. \ Further, it is usually of interest to study the (joint) 
\textbf{distribution} of one (or more) \textbf{variable}(s) and/or \textbf{%
measurement}(s) to be taken on individual members of the population. This
distribution often characterizes the process and population in a way that is
of interest to the researcher. The measurement(s) taken on an individual is
sometimes referred to as an observation on the individual. Although the
collection of observations on the individuals of a population are also often
referred to as the population (c.f. \cite{F58}), for clarity, we shall
carefully draw a distinction between the individuals and the their
associated measurement(s).

To emphasize again the importance of regarding both hypothetical and actual
individuals as the population of interest, we present an example where the
population consists \textit{entirely} of conceptual individuals:

\begin{example}
Suppose a researcher is interested in comparing treatments A and B. The
population of interest is adult women. Presently, no adult woman is
receiving treatment A or treatment B. Associated with treatment A is the
hypothetical population of adult women who would be receiving treatment A.
There is also a similar hypothetical population of adult women who would be
receiving treatment B. At the beginning of the study, both of these
populations contain only conceptual individuals. The representative sample
is empty. So how does the researcher obtain samples from these two
populations? Simply  take a sample from the representative sample of adult
women. Split this sample into two subsamples. The subsample that receives
treatment A is a sample from the population of adult women who would be
receiving treatment A. Similarly, for treatment B. In both cases, the
researcher's sample is the representative sample. As strange as it may seem,
each of the researcher's samples constitute a census.
\end{example}

Let us consider another example.

\begin{example}
A pollster is interested in the proportion of individuals who will say
they will vote for a given presidential candidate one month before the
actual election. At that point in time there will be a finite number of
voters. So how does one view this collection as a representative sample of
what some process has produced and not the population of interest? What if
the election to be studied by the pollster is to occur four years from now?
The collection of voters that will exist one month before the election date
has yet to be (completely) generated. Consequently, this collection of
voters can only be viewed as a representative sample of what the process can
generate one month before the election as well as at any time. It then
follows that a \textbf{census} occurs when the measurement(s) of interest is
taken on each individual in the representative sample.
\end{example}

\citet[p. 33]{F58}
states: \textquotedblleft \lbrack w]hen a large number of individuals are
measured in respect of physical dimensions, weight, colour, density, etc.,
it is possible to describe with some accuracy the \textit{population} of
which our experience may be regarded as a sample. By this means it may be
possible to distinguish it from other populations differing in their genetic
origin, or in environmental circumstances.\textquotedblright\ Further, 
\citet[p. 41]{F58}
states the following about population and frequency distribution.
\textquotedblleft The idea of an infinite \textbf{population} distributed in
a \textbf{frequency distribution} in respect of one or more characters is
fundamental to all statistical work. From a limited experience, for example,
of individuals of a species, or of the weather of a locality, we may obtain
some idea of the infinite hypothetical population from which our sample is
drawn, and so of the probable nature of future samples to which our
conclusions are to be applied.\textquotedblright\ Here, he alludes to a
process (genetic origin or environmental circumstances), describes a
population as an infinite collection, and does not make a distinction
between the population as a collection of individuals or as the collection
of the measurements on these individuals.

\cite{GHM07} state:

\begin{quotation}
Statistical analysis involves use of observational data together with domain
knowledge to develop a model to study and understand a data-generating
process. The data analysis is used to refine the model or possibly to select
a different model, to determine appropriate values for terms in the model,
and to use the model to make inferences concerning the process. This has
been the paradigm followed by statisticians for centuries. The advances in
statistical theory over the past two centuries have not changed the
paradigm, but they have improved the specific methods. Not only has the
exponentially-increasing computational power allowed use of more detailed
and better models, however, it has shifted the paradigm slightly. Many
alternative views of the data can be examined. Many different models can be
explored. Massive amounts of simulated data can be used to study the
model/data possibilities.
\end{quotation}

The process must first generate an individual on which one or more
measurements are to be taken resulting in the generation of data. We
reiterate our definition of population as the collection of individuals the
process has or could have generated.

W. E. Deming devoloped sampling techniques that are still used today by the
United States Department of the Census and the Bureau of Labor Statistics.
He is perhaps best known and recognized by the Japanese for his post-World
War II work in helping the Japanese capture world markets in several key
industries such as automotive, steel, and electronics. Their success was due
in large part to producing high quality products. \cite{D53} stated the
following:

\begin{quotation}
Statistical data are supposedly collected to provide a rational basis for
action. The action may call for the enumerative interpretation of the data,
or it may call for the analytic interpretation. The aim here is to exhibit
some of the consequences of failing to distinguish between the enumerative
and the analytic uses of data. This distinction is necessary in the
statement of the aims of a survey, census, or experiment, in order that the
plans for the collection of the data and for the tabulations may most
economically meet the needs of the consumer, and it is equally important in
the interpretation of data. Thus, to draw on a result from a later
paragraph, information obtained in a complete census concerning every person
in an area (e.g., on occupation, income, or education) still possesses for
analytic purposes a sampling error that is actually about a quarter as great
as the sampling error of a 6 per cent sample. The consequences are
far-reaching. In using a census-table for analytic purposes, even though the
figures come from a perfect complete count, it is therefore necessary to
bear in mind that small numbers in a cell are unreliable in the sense that
they have a standard error, just as if they had arisen in sampling, as
indeed they did. Moreover, in the planning of a complete census, it is
therefore imperative to use sampling for every bit of information that is
not necessary as an aid to complete coverage, or required to give detail for
small areas (such as the block statistics). Name, relationship to the head,
age, sex, marital status, color are probably all necessary for the sake of
completeness of coverage. These things, plus a few questions on rent,
tenure, year built, will provide the information required for the block
statistics.
\end{quotation}

Thus Deming pointed out that \textit{even a census} is a sample. Recall that
a census occurs when the measurement(s) of interest is taken on each
individual in the representative sample. Although Deming does not explicitly
point to an infinite population, the implication here is that the population
is a collection of individuals that contains more than the individuals that
have been generated by a process.

Through out most of their textbook, 
\citet{MNF01}
refer to what we call the representative sample as the population. This can
be summarized from the statement they make on page 376: \textquotedblleft
The condition that the population is large relative to the size of the
sample will be satisfied if the population is, say, at least 20 times as
large.\textquotedblright\ However, in Chapter 31 titled \textquotedblleft
Statistical Process Control,\textquotedblright\ they state, \textquotedblleft
[w]e can accommodate processes in our sample-versus-population framework:
think of the population as containing all the outputs that would be produced
by the process if it ran forever in its present state. The outputs produced
today or this week are a sample from this population. Because the population
doesn't actually exist now, it is simpler to speak of a process and of
recent output as a sample from the process in its present
state.\textquotedblright\ Note, however, that we maintain that this
viewpoint is applicable and appropriate to all processes and their
corresponding populations, not just to the context of statistical process
control in industrial settings.

\section{Describing the Data Model}

\citet[p. 292, Ex. 7.1]{M01}
gives an example of a process packaging frozen orange juice concentrate in
six ounce cardboard containers. One measurement of interest is whether the
container leaks or not. A number that completely characterizes the
distribution of this measurement is the rate, $p$, at which the process is
producing containers that leak. Another equivalent view of the meaning of $p$
is that $p$ is the probability that a given container will leak. Since the
number $p$ characterizes the population of interest, it is generally
referred to as a \textbf{parameter}. If $N$ of these containers of frozen
orange juice concentrate\ produced by this process are to be purchased by a
store, the $N$ items only constitute a representative sample from the
process. There is no reason for the store to view this collection of $N$
items as a population. The uncertainty as to which collection of $N$ cans
the process has generated must be taken into account. The store may be
interested in the number $Y_{N}$ or proportion $\widehat{p}_{N}$ of these $N$
containers that leak, but these values will depend on the sample received.
Such numbers that characterize a sample (whether a representative sample or
researcher's sample) are referred to as \textbf{statistics}. As statistics,
they have distributions often referred to as \emph{sampling distributions}.
Their distributions take into account the uncertainty as to which
representative sample the process has generated with respect to the
measurement of interest on the individuals.

It is useful to classify variables and measurements of interest. Virtually
all of the variables we will study can be classified as either discrete or
continuous. A \textbf{discrete} variable is a variable in which its set of
possible values is a countable (i.e. either a finite or an uncountably
infinite) set. Measurements such as length, weight, time, area, volume, and
rates are variables in which the set of possible outcomes is a continuum of
the real numbers, typically, some interval of the real numbers. These are
examples of \textbf{continuous} variables. In the orange juice concentrate
example, the measurement of interest has only two possible outcome -- leaks
or does not leak, and is therefore a discrete measurement. \ 

Often the process and population of interest is studied by examining the
(joint) distribution of the measurement(s) of interest. A useful way to view
the distribution of a continuous variable is with a \textbf{probability
density curve} (function). For many continuous variables $X$ of interest,
the density (function) curve describing its distribution is simply a
(function) curve $y=f\left( x\right) $ such that the curve is on or above
the measurement axis and the total area under the curve is one. Further,
area under the curve and over an interval is the proportion of the
population having their measurements $X$ falling in this interval. One can
view the probability density function as a continuous analog of the
population proportion. We note for completeness that there exist continuous
measurements that do not have a probability density function that describes
its distribution, e.g., the Cantor distribution 
\citep[cf.][pp. 40 ff.]{G05}%
. However, this is not the case for the vast majority of measurements taken
by researchers.

It is of importance to keep in mind that the researcher is typically
interested in making an inference about the process and/or the associated
population. The population is an uncountable collection and for this reason
alone it is impossible for the researcher view the entire population. Since
one cannot view the whole population, then only partial information about
the process and population can be obtained by looking at a finite
subcollection (sample) of what the process has produced. Ideally, we would
like for this sample to be representative of the population. What the
process \textit{has} produced is a representative sample of what the process 
\textit{can} produce. Often this sample is either too large or does not
initially exist, so that it would be impossible (or at least impractical)
for the researcher to view the full representative sample. One solution is
to take a \textquotedblleft representative\textquotedblright\ sample from
this representative sample. Unfortunately, there is no sampling method that
guarantees a representative sample. The best for which one can hope is a
method that will \textquotedblleft most likely\textquotedblright\ produce a
representative sample. Also, it should be kept in mind that even after a
sample is produced by the process, or selected by the researcher (as in the
case of a designed experiment), and before the sample is examined, the
uncertainty about \textit{which} representative sample has been obtained is
the same as that which existed \textit{before} the sample was brought into
existence. It is this uncertainty that must be accounted for by the
researcher in making an inference. The often used arguement that this sample
generated by the process is now \textquotedblleft fixed\textquotedblright\
which is to imply that somehow this uncertainty has been removed. \ But
consider the following scenario: a penny is to be flipped and a
\textquotedblleft head\textquotedblright\ or \textquotedblleft
tail\textquotedblright\ is to be observed. Assume the model that the
probability of a \textquotedblleft head\textquotedblright\ is $50\%$. If the
penny has been flipped and the face-up side has not been observed, the
outcome is \textquotedblleft fixed\textquotedblright\ but the uncertainty as
to whether the outcome is a \textquotedblleft head\textquotedblright\ still
remains to be $50\%$.

\section{Further Examples}

\begin{example}
\bigskip A coin is to be flipped. A measurement $X$ to be taken on the coin
has the value $1$ if the face-up side of the coin is a \textquotedblleft
head\textquotedblright\ and a value of $0$ if the face-up side of the coin
is a \textquotedblleft tail.\textquotedblright\ The process is the flipping
of the coin. The population is the collection of all possible flips of the
coin. This is an uncountable collection of possibilities. For the American
penny, it is common to assume that the outcomes of a \textquotedblleft
head\textquotedblright\ ($X=1$) and a \textquotedblleft
tail\textquotedblright\ ($X=0$) are equally likely (fifty-fifty). Assume
this model. Suppose that the coin has been flipped but the outcome of
\textquotedblleft head\textquotedblright\ or \textquotedblleft
tail\textquotedblright\ has not been revealed. Does the knowledge that the
coin has been flipped provide information that would cause one to now assume
a different model than the fifty-fifty model? No, since we do not have a
reason to assume our level of uncertainty has changed.
\end{example}

\begin{example}
A penny is to be placed on a flat surface, spun, and its resting side that
is up is to be observed. This is a process of generating \textquotedblleft
heads\textquotedblright\ and \textquotedblleft tails.\textquotedblright\ The
collection of all possible ways of spinning this penny is an uncountable
collection. Let $p$ be the rate at which the process is generating spins
that result in a resting side up of \textquotedblleft
heads.\textquotedblright\ Define the measurement $X$ to have the value of $1$
if the spin of the penny results in the coin resting heads-up and $X=0$ if
tails-up. The coin is now spun two hundred times and it is observed that the
number of times the coin was \textquotedblleft heads\textquotedblright\ in
its resting position was $83$. Thus, the proportion of the $200$ spins that
resulted in \textquotedblleft heads\textquotedblright\ is $83/200=0.415$. Is 
$p=0.415$ or is $0.415$ an estimate of the parameter $p$?
\end{example}

\begin{example}
A process has generated a collection of healthy adults. The population of
interest is the collection of healthy adults the process has or could have
generated. At three hour time intervals, the body temperatures $%
X_{1},X_{2},\ldots ,X_{8}$ of each adult are to be measured over a 24-hour
period and the average%
\begin{equation*}
\overline{X}=\frac{X_{1}+X_{2}+\ldots +X_{8}}{8}
\end{equation*}%
is recorded. The measurement $\overline{X}$ (daily average body temperature)
on each health adult is a continuous variable. Hence, there is a density
curve that describes how $\overline{X}$ is distributed over the population
of healthy adults. It is generally assumed that average body temperature $%
\mu $, in degrees Fahrenheit, is $98.6^{\mathbf{o}}$ F. This is the mean of
the distribution of body temperature measurements $X$ over all healthy
adults. The assumption that $\mu =98.6^{\mathbf{o}}$ F is a model. It can be
shown that under the model $\mu =98.6^{\mathbf{o}}$ F, the average $\mu _{%
\overline{X}}$ over all healthy adults is also $98.6^{\mathbf{o}}$ F.
\end{example}

\bigskip

\section{Some Distributional Results}

Let the $X$ measurements on the $N$ individuals in the representative sample
be denoted by $X_{1},\ldots ,X_{N}$. \ From these $N$ individuals, the
researcher will take a subcollection of size $n$. Without loss of generality
let us re-number the measurements $X_{1},\ldots ,X_{N}$ so that the $X$
measurements on the $n$ individuals that appear in the researcher's sample
are $X_{1},\ldots ,X_{n}$ and the $X$ measurements on the $N-n$ individuals
not selected by the researcher are $X_{n+1},\ldots ,X_{N}$. We will make the
assumption that the $X$ measurements $X_{1},\ldots ,X_{N}$ are independent
and identially distributed. It follows that the $X$ measurements of any
subcollection of these $N$ individual that are selected at random is a
random sample. Hence, the researcher's sample is a random sample provided
the representative sample is a random sample. Assume the common distribution
is a Bernoulli distribution with parameter $p$ such that $0<p<1$.

We define the following statistics:%
\begin{eqnarray*}
Y_{N} &=&X_{1}+\ldots +X_{N}\text{, }\widehat{p}_{N}=\frac{1}{N}Y_{N}\text{,}
\\
Y_{n} &=&X_{1}+\ldots +X_{n}\text{, }\widehat{p}_{n}=\frac{1}{n}Y_{n}\text{,}
\\
Y_{N-n} &=&Y_{N}-Y_{n}\text{, and }\widehat{p}_{N-n}=\frac{1}{N-n}Y_{N-n}%
\text{.}
\end{eqnarray*}%
Under our model, the statistic $Y_{N}$ has a Binomial distribution with
distributional parameters $N$ and $p$. The conditional distribution of $%
Y_{n} $ given $\mathbf{X}$ is a hypergeometric distribution with
distributional parameters $Y_{N}$ and $N$, where $\mathbf{X}=\left[
X_{1},\ldots ,X_{N}\right] ^{\mathbf{T}}$. It is now easy to show that%
\begin{eqnarray*}
E\left( Y_{n}\left\vert \mathbf{X}\right. \right) &=&n\frac{Y_{N}}{N}=n%
\widehat{p}_{N}\text{, }V\left( Y_{n}\left\vert \mathbf{X}\right. \right)
=\left( 1-\frac{n-1}{N-1}\right) n\widehat{p}_{N}\left( 1-\widehat{p}%
_{N}\right) \text{,} \\
E\left( \widehat{p}_{n}\left\vert \mathbf{X}\right. \right) &=&n\frac{Y_{N}}{%
N}=\widehat{p}_{N}\text{, and }V\left( \widehat{p}_{n}\left\vert \mathbf{X}%
\right. \right) =\left( 1-\frac{n-1}{N-1}\right) \frac{\widehat{p}_{N}\left(
1-\widehat{p}_{N}\right) }{n}\text{.}
\end{eqnarray*}%
Conditioned on $\mathbf{X}$, the statistic $\widehat{p}_{n}$ is an unbiased
predictor of $\widehat{p}_{N}$, and as $n$ approaches $N$, it becomes a more
precise predictor of $\widehat{p}_{N}$. Further, one can show that the
unconditional distribution of $Y_{n}$ has a Binomial distribution with
distributional parameters $n$ and $p$ with the unconditional distribution of 
$\widehat{p}_{n}$ given by%
\begin{equation*}
P\left( \widehat{p}_{n}=\frac{y}{n}\right) =P\left( Y_{n}=y\right) \text{.}
\end{equation*}%
Hence,%
\begin{equation*}
E\left( \widehat{p}_{n}\right) =p\text{ and }V\left( \widehat{p}_{n}\right) =%
\frac{p\left( 1-p\right) }{n}\text{.}
\end{equation*}%
Thus, the researcher's sample proportion $\widehat{p}_{n}$ is an unbiased
estimator of the population proportion $p$, and as $n$ increases it is a
more precise estimator of $p$.

An estimator for $V\left( \widehat{p}_{n}\right) $ is%
\begin{equation*}
\widetilde{V}\left( \widehat{p}_{n}\right) =\frac{\widehat{p}_{n}\left( 1-%
\widehat{p}_{n}\right) }{n}\text{.}
\end{equation*}%
However,%
\begin{eqnarray*}
E\left( \frac{\widehat{p}_{n}\left( 1-\widehat{p}_{n}\right) }{n}\right) &=&%
\frac{E\left( \widehat{p}_{n}\right) -E\left( \widehat{p}_{n}^{2}\right) }{n}%
=\frac{E\left( \widehat{p}_{n}\right) -V\left( \widehat{p}_{n}\right) -\left[
E\left( \widehat{p}_{n}\right) \right] ^{2}}{n} \\
&=&\frac{p-\frac{p\left( 1-p\right) }{n}-p^{2}}{n}=\frac{n-1}{n}\frac{%
p\left( 1-p\right) }{n}\text{.}
\end{eqnarray*}%
Hence, $\widetilde{V}\left( \widehat{p}_{n}\right) $ is a biased estimator
of $V\left( \widehat{p}_{n}\right) $. We can now see that%
\begin{equation*}
\widehat{V}\left( \widehat{p}_{n}\right) =\frac{n}{n-1}\frac{\widehat{p}%
_{n}\left( 1-\widehat{p}_{n}\right) }{n}=\frac{\widehat{p}_{n}\left( 1-%
\widehat{p}_{n}\right) }{n-1}
\end{equation*}%
is an unbiased estimator of $V\left( \widehat{p}_{n}\right) $. \ This
results is given in 
\citet{SMOG12}%
.

For $n<N$ and $\lambda =N/n$, if $n\rightarrow \infty $, then $%
N-n\rightarrow \infty $. We observe that%
\begin{equation*}
\frac{\widehat{p}_{n}-p}{\sqrt{\frac{p\left( 1-p\right) }{n}}}\overset{d}{%
\longrightarrow }N\left( 0,1\right) \text{ and }\frac{\widehat{p}_{N-n}-p}{%
\sqrt{\frac{p\left( 1-p\right) }{N-n}}}\overset{d}{\longrightarrow }N\left(
0,1\right)
\end{equation*}%
according to the central limit theorem as $n\rightarrow \infty $. \ Another
limiting distribution result states that%
\begin{equation*}
\frac{\widehat{p}_{n}-p}{\sqrt{\frac{\widehat{p}_{n}\left( 1-\widehat{p}%
_{n}\right) }{n}}}\overset{d}{\longrightarrow }N\left( 0,1\right) \text{.}
\end{equation*}%
as $n\rightarrow \infty $. \ See 
\citet[p. 249, Example 7.7.2]{BE92}%
. We now examine the limiting distributions of%
\begin{equation*}
\frac{\widehat{p}_{n}-\widehat{p}_{N}}{\sqrt{\frac{N-n}{N}}\sqrt{\frac{%
p\left( 1-p\right) }{n}}}\text{, }\frac{\widehat{p}_{n}-\widehat{p}_{N}}{%
\sqrt{\frac{N-n}{N}}\sqrt{\frac{\widehat{p}_{n}\left( 1-\widehat{p}%
_{n}\right) }{n}}}\text{, and }\frac{\widehat{p}_{n}-\widehat{p}_{N}}{\sqrt{%
\frac{N-n}{N}}\sqrt{\frac{\widehat{p}_{N}\left( 1-\widehat{p}_{N}\right) }{N}%
}}\text{.}
\end{equation*}

Observe that we can write%
\begin{eqnarray*}
\frac{\widehat{p}_{n}-\widehat{p}_{N}}{\sqrt{\frac{N-n}{N}}\sqrt{\frac{%
p\left( 1-p\right) }{n}}} &=&\frac{\frac{N-n}{N}\left( \widehat{p}_{n}-%
\widehat{p}_{N-n}\right) }{\sqrt{\frac{N-n}{N}}\sqrt{\frac{p\left(
1-p\right) }{n}}} \\
&=&\sqrt{\frac{N-n}{N}}\frac{\widehat{p}_{n}-p}{\sqrt{\frac{p\left(
1-p\right) }{n}}}-\sqrt{\frac{n}{N}}\frac{\widehat{p}_{N-n}-p}{\sqrt{\frac{%
p\left( 1-p\right) }{N-n}}} \\
&=&\sqrt{\frac{N-n}{N}}Z_{n}-\sqrt{\frac{n}{N}}Z_{N-n}\text{,}
\end{eqnarray*}%
where%
\begin{equation*}
Z_{n}=\frac{\widehat{p}_{n}-p}{\sqrt{\frac{p\left( 1-p\right) }{n}}}\text{
and }Z_{N-n}=\frac{\widehat{p}_{N-n}-p}{\sqrt{\frac{p\left( 1-p\right) }{N-n}%
}}\text{.}
\end{equation*}%
We see that%
\begin{eqnarray*}
\sqrt{\frac{N-n}{N}} &=&\sqrt{\frac{\lambda n-n}{\lambda n}}=\sqrt{\frac{%
\lambda -1}{\lambda }}\text{ and} \\
\sqrt{\frac{n}{N}} &=&\sqrt{\frac{n}{\lambda n}}=\sqrt{\frac{1}{\lambda }}%
\text{.}
\end{eqnarray*}%
Using Slutsky's theorem, we have%
\begin{equation*}
U_{n}=\sqrt{\frac{\lambda -1}{\lambda }}Z_{n}\overset{d}{\longrightarrow }%
N\left( 0,\frac{\lambda -1}{\lambda }\right) \text{ and }V_{n}=\sqrt{\frac{1%
}{\lambda }}Z_{\left( \lambda -1\right) n}\overset{d}{\longrightarrow }%
N\left( 0,\frac{1}{\lambda }\right)
\end{equation*}%
as $n\rightarrow \infty $. Making the transformation%
\begin{equation*}
T_{n}=U_{n}-V_{n}\text{ and }Q_{n}=V_{n}
\end{equation*}%
with inverse transformation $U_{n}=T_{n}+Q_{n}$ and $V_{n}=Q_{n}$ with
Jacobian $J=1$, we have%
\begin{equation*}
f_{T_{n},Q_{n}}\left( t,q\right) =f_{U_{n}}\left( t+q\right) f_{V_{n}}\left(
q\right) \text{.}
\end{equation*}%
It follows that%
\begin{eqnarray*}
\lim_{n\rightarrow \infty }f_{T_{n},Q_{n}}\left( t,q\right)
&=&\lim_{n\rightarrow \infty }f_{U_{n}}\left( t+q\right) f_{V_{n}}\left(
q\right) =\lim_{n,\rightarrow \infty }f_{U_{n}}\left( t+q\right)
\lim_{n\rightarrow \infty }f_{V_{n}}\left( q\right) \\
&=&f_{N\left( 0,\frac{\lambda -1}{\lambda }\right) }\left( t+q\right)
f_{N\left( 0,\frac{1}{\lambda }\right) }\left( q\right) \text{,}
\end{eqnarray*}%
where $f_{N\left( 0,\sigma ^{2}\right) }\left( y\right) $ is the probability
density function of a Normal distribution with mean $0$ and variance $\sigma
^{2}$. We then have%
\begin{equation*}
\lim_{n\rightarrow \infty }f_{T_{n}}\left( t\right) =\int\nolimits_{-\infty
}^{\infty }f_{N\left( 0,\frac{\lambda -1}{\lambda }\right) }\left(
t+q\right) f_{N\left( 0,\frac{1}{\lambda }\right) }\left( q\right) dq=\frac{1%
}{\sqrt{2\pi }}e^{-t^{2}/2}\text{.}
\end{equation*}%
Hence,%
\begin{equation*}
\frac{\widehat{p}_{n}-\widehat{p}_{N}}{\sqrt{\frac{N-n}{N}}\sqrt{\frac{%
p\left( 1-p\right) }{n}}}\overset{d}{\longrightarrow }N\left( 0,1\right) 
\text{.}
\end{equation*}%
Again using Slutsky's theorem, we have%
\begin{equation*}
\frac{\widehat{p}_{n}-\widehat{p}_{N}}{\sqrt{\frac{N-n}{N}}\sqrt{\frac{%
\widehat{p}_{n}\left( 1-\widehat{p}_{n}\right) }{n}}}\overset{d}{%
\longrightarrow }N\left( 0,1\right) \text{ and }\frac{\widehat{p}_{n}-%
\widehat{p}_{N}}{\sqrt{\frac{N-n}{N}}\sqrt{\frac{\widehat{p}_{N}\left( 1-%
\widehat{p}_{N}\right) }{n}}}\overset{d}{\longrightarrow }N\left( 0,1\right)
\end{equation*}%
as $n\rightarrow \infty $. Using similar arguements, one can show that%
\begin{equation*}
\frac{\overline{X}_{n}-\overline{X}_{N}}{\sqrt{\frac{N-n}{N}}\frac{S_{n}}{%
\sqrt{n}}}\overset{d}{\longrightarrow }N\left( 0,1\right) \text{,}
\end{equation*}%
where $\overline{X}_{n}$ and $\overline{X}_{N}$ are the means of the
researcher's sample and the representative sample and $S_{n}$ is the
standard deviation of the researcher's sample.

\section{Confidence and Prediction Intervals}

\subsection{Confidence intervals for the population proportion $p$}

Suppose that a process is producing individuals that possess a
characteristic of interest at a rate $p$ and that a researcher will have a
sample of $n$ individuals whose $X$ measurements are to be taken. \ We
assume here that $0<p<1$. \ The random variable $X$ will assume the value $1$
if the individual possesses the characteristic of interest and $0$ if it
does not. \ The random variable $X$ thus has a Bernoulli distribution with
distributional parameter $p$. \ Virtually every elementary statistics
textbook introduces the Wald approximate confidence interval for $p$, based
on the observation that for sufficiently large $n$, $Y_{n}\overset{%
\centerdot }{\sim }N(n\widehat{p}_{n},n\widehat{p}_{n}(1-\widehat{p}_{n}))$
, which leads to the approximate $100(1-\alpha )\%$ confidence interval%
\begin{equation*}
\left( \widehat{p}_{n}-z_{\alpha /2}\sqrt{\frac{\widehat{p}_{n}(1-\widehat{p}%
_{n})}{n}},\widehat{p}_{n}+z_{\alpha /2}\sqrt{\frac{\widehat{p}_{n}(1-%
\widehat{p}_{n})}{n}}\right) \text{.}
\end{equation*}%
for $p$. This is based on the random variable $\left( \widehat{p}%
_{n}-p\right) /\sqrt{\widehat{p}_{n}(1-\widehat{p}_{n})/n}$ having an
approximate standard normal distribution. Another \textquotedblleft
large\textquotedblright\ sample confidence interval for $p$ is%
\begin{equation*}
\left( \frac{2n\widehat{p}_{n}+z_{\alpha /2}^{2}-z_{\alpha /2}\sqrt{%
z_{\alpha /2}^{2}+4n\widehat{p}_{n}\left( 1-\widehat{p}_{n}\right) }}{%
2\left( n+z_{\alpha /2}^{2}\right) },\frac{2n\widehat{p}_{n}+z_{\alpha
/2}^{2}+z_{\alpha /2}\sqrt{z_{\alpha /2}^{2}+4n\widehat{p}_{n}\left( 1-%
\widehat{p}_{n}\right) }}{2\left( n+z_{\alpha /2}^{2}\right) }\right) \text{.%
}
\end{equation*}%
This is based on the random variable $\left( \widehat{p}_{n}-p\right) /\sqrt{%
p(1-p)/n}$ having an approximate standard Normal distribution. \ Several
\textquotedblleft exact\textquotedblright\ confidence intervals for $p$,
based on $Y_{n}$ having a binomial distribution with distribution parameters 
$n$ and $p,$ are known in the literature and implemented in statistical
software. The most well known of these was given by \cite{CP34}. \ For a
discussion of other \textquotedblleft exact\textquotedblright\ confidence
intervals for $p$, see 
\citet[p. 821]{T14}%
.

\bigskip

\subsection{Prediction intervals for the representative sample proportion $%
\widehat{p}_{N}$}

A corresponding approximate $100(1-\alpha )\%$ prediction interval for $%
\widehat{p}_{N}$ is given by

\begin{equation*}
\left( \widehat{p}_{n}-z_{\alpha /2}\sqrt{1-\frac{n}{N}}\sqrt{\frac{\widehat{%
p}_{n}(1-\widehat{p}_{n})}{n}},\widehat{p}_{n}+z_{\alpha /2}\sqrt{1-\frac{n}{%
N}}\sqrt{\frac{\widehat{p}_{n}(1-\widehat{p}_{n})}{n}}\right)
\end{equation*}%
for \textquotedblleft large\textquotedblright\ $n$. This is based on the
limiting distribution of the statistic%
\begin{equation*}
\frac{\widehat{p}_{n}-\widehat{p}_{N}}{\sqrt{\frac{N-n}{N}}\sqrt{\frac{%
\widehat{p}_{n}\left( 1-\widehat{p}_{n}\right) }{n}}}
\end{equation*}%
being a standard Normal distribution. Since $\sqrt{1-n/N}<1$ and $N$ is not
known, an approximate at least $100(1-\alpha )\%$ prediction interval for $%
\widehat{p}_{N}$ is%
\begin{equation*}
\left( \widehat{p}_{n}-z_{\alpha /2}\sqrt{\frac{\widehat{p}_{n}(1-\widehat{p}%
_{n})}{n}},\widehat{p}_{n}+z_{\alpha /2}\sqrt{\frac{\widehat{p}_{n}(1-%
\widehat{p}_{n})}{n}}\right)
\end{equation*}%
for $N$ \textquotedblleft large\textquotedblright\ relative to $n$. Also,
the observed value of the random interval,%
\begin{eqnarray*}
&&%
\left(%
\frac{2\widehat{p}_{n}+z_{\alpha /2}^{2}\frac{N-n}{nN}}{2\left( 1+z_{\alpha
/2}^{2}\frac{N-n}{nN}\right) }-\frac{\sqrt{\left( 2\widehat{p}_{n}+z_{\alpha
/2}^{2}\frac{N-n}{nN}\right) ^{2}-4\left( 1+z_{\alpha /2}^{2}\frac{N-n}{nN}%
\right) \widehat{p}_{n}^{2}}}{2\left( 1+z_{\alpha /2}^{2}\frac{N-n}{nN}%
\right) },%
\right.
\\
&&%
\qquad\qquad\qquad%
\left.%
\frac{2\widehat{p}_{n}+z_{\alpha /2}^{2}\frac{N-n}{nN}}{2\left( 1+z_{\alpha
/2}^{2}\frac{N-n}{nN}\right) }+\frac{\sqrt{\left( 2\widehat{p}_{n}+z_{\alpha
/2}^{2}\frac{N-n}{nN}\right) ^{2}-4\left( 1+z_{\alpha /2}^{2}\frac{N-n}{nN}%
\right) \widehat{p}_{n}^{2}}}{2\left( 1+z_{\alpha /2}^{2}\frac{N-n}{nN}%
\right) }%
\right)%
\end{eqnarray*}%
is an approximate $100\left( 1-\alpha \right) \%$ prediction interval for $%
\widehat{p}_{N}$ which is based on the limiting distribution of the statistic%
\begin{equation*}
\frac{\widehat{p}_{n}-\widehat{p}_{N}}{\sqrt{\frac{N-n}{N}}\sqrt{\frac{%
\widehat{p}_{N}\left( 1-\widehat{p}_{N}\right) }{n}}}
\end{equation*}%
being a standard Normal distribution. Note that%
\begin{equation*}
\frac{N-n}{nN}=\frac{1}{n}\left( 1-\frac{n}{N}\right) <\frac{1}{n}\text{.}
\end{equation*}%
In the aforementioned prediction interval for $\widehat{p}_{N}$, replacing $%
\left( N-n\right) /\left( nN\right) $ with $1/n$ when $N$ is
\textquotedblleft large\textquotedblright\ relative to $n$, we obtain the
prediction interval%
\begin{eqnarray*}
&&%
\left(%
\frac{2\widehat{p}_{n}+z_{\alpha /2}^{2}/n}{2\left( 1+z_{\alpha
/2}^{2}/n\right) }-\frac{\sqrt{\left( 2\widehat{p}_{n}+z_{\alpha
/2}^{2}/n\right) ^{2}-4\left( 1+z_{\alpha /2}^{2}/n\right) \widehat{p}%
_{n}^{2}}}{2\left( 1+z_{\alpha /2}^{2}/n\right) },%
\right.
\\
&&%
\qquad\qquad\qquad%
\left.%
\frac{2\widehat{p}_{n}+z_{\alpha /2}^{2}/n}{2\left( 1+z_{\alpha
/2}^{2}/n\right) }+\frac{\sqrt{\left( 2\widehat{p}_{n}+z_{\alpha
/2}^{2}/n\right) ^{2}-4\left( 1+z_{\alpha /2}^{2}/n\right) \widehat{p}%
_{n}^{2}}}{2\left( 1+z_{\alpha /2}^{2}/n\right) }%
\right)%
\end{eqnarray*}%
is an approximate at least $100\left( 1-\alpha \right) \%$ prediction
interval for $\widehat{p}_{N}$ that does not depend of the knowing the value
of $N$.

Thus we see that the prediction interval for a statistic describing the
representative sample is essentially the confidence interval for the
corresponding population parameter with the extra factor $\sqrt{(N-n)/N}$,
which is sometimes called the \textquotedblleft finite population
correction\textquotedblright\ factor, as in 
\citet[p. 24]{C77}%
. But under the view for which we are advocating, there are no finite
populations, so what has been thought of as a \textquotedblleft confidence
interval for a parameter of a finite population\textquotedblright\ is really
a prediction interval for a statistic of the representative sample in what
Deming called an enumerative study.

\bigskip

\section{Conclusion}

We make the case that researchers are interested in drawing inferences about
a process and population, and that a population is inherently an uncountably
infinite collection. \ That which is often referred to as a
\textquotedblleft finite population\textquotedblright\ is better termed the
\textquotedblleft representative sample\textquotedblright\ because it is
representative of what the process can produce, and is a sample in the sense
that that it possesses uncertainty. \ To distiguish the set of individuals
actually produced by the process from the sample collected for study by the
researcher, we call the latter the \textquotedblleft researcher's
sample.\textquotedblright\ \ Finally, we draw a clear distinction between
estimation and prediction, where the former refers to inference drawn about
the population and the latter about the collection of all actual individuals
produced by the process, herein called the \textquotedblleft representative
sample.\textquotedblright\ \ Various confidence and prediction intervals
were derived and compared.

\section{Disclosure}
No potential competing interest was reported by the authors.

\bigskip

\bigskip 
\bibliographystyle{natbib-harv}
%

\end{document}